\documentclass[10pt,twocolumn,twoside]{IEEEtran} 

\usepackage{amsfonts}
\usepackage{amssymb}
\usepackage{amsmath}
\usepackage{amsthm}
\usepackage{graphicx}
\usepackage{subfigure}
\usepackage{multirow}
\usepackage[verbose,nospace,sort]{cite}
\usepackage{algpseudocode}
\usepackage{algorithm}

\begin{document}
\newcommand{\xB}{\mathbf{B}}
\newcommand{\xE}{\text{E}}
\newcommand{\xe}{\mathbf{e}}
\newcommand{\xGF}{\text{GF}}
\newcommand{\xtr}{\text{tr}}
\newcommand{\xd}{\mathbf{d}}
\newcommand{\xH}{\mathbf{H}}
\newcommand{\xI}{\mathbf{I}}
\newcommand{\xtI}{\text{I}}
\newcommand{\xtlog}{\text{log}}
\newcommand{\xmkl}{_{k,l}}
\newcommand{\xmkj}{_{k,j}}
\newcommand{\xQ}{\mathbf{Q}}
\newcommand{\xP}{\mathbf{P}}
\newcommand{\xp}{\mathbf{p}}
\newcommand{\xR}{\mathbf{R}}
\newcommand{\xSF}{\text{SF}}
\newcommand{\xtSINR}{\text{SINR}}
\newcommand{\xu}{\mathbf{u}}
\newcommand{\xU}{\mathbf{U}}
\newcommand{\xv}{\mathbf{v}}
\newcommand{\xtv}{\text{v}}
\newcommand{\xV}{\mathbf{V}}
\newcommand{\xy}{\mathbf{y}}
\newcommand{\xX}{\mathbf{X}}
\newcommand{\xx}{\mathbf{x}}
\newcommand{\xz}{\mathbf{z}}

\title{Sub-Stream Fairness and Numerical Correctness in MIMO Interference Channels}

\author{Cenk M. Yetis, \IEEEmembership{Member, IEEE,
} Yong Zeng, \IEEEmembership{Student Member, IEEE, } Kushal Anand, \IEEEmembership{Student Member, IEEE, }
\\ Yong Liang Guan, \IEEEmembership{Member, IEEE} and Erry Gunawan, \IEEEmembership{Member, IEEE}
\thanks{
Part of the material in this paper was presented in the Proceedings of IEEE ISWTA 2013.}
\thanks{The work of C. M. Yetis in part and Y. Zeng, K. Anand, Y. L. Guan, and E. Gunawan in full was supported by the Department of the Navy Grant N62909-12-1-7015 issued by Office of Naval Research Global.}
\thanks{C. M. Yetis was with the INFINITUS Research Lab, Nanyang Technological University, Singapore 639798. He is now with the Department of Electrical and Electronics Engineering, Mevlana University, Selcuklu, Konya, 42003, Turkey (e-mail:cenkmyetis@yahoo.com).}
\thanks{Y. Zeng, K. Anand, Y. L. Guan and E. Gunawan are with the School of Electrical and Electronic Engineering, Nanyang Technological University, Singapore 639798 (e-mail:\{ze0003ng, kush0004\}@e.ntu.edu.sg, \{eylguan, egunawan\}@ntu.edu.sg).}
}

\maketitle
\begin{abstract}
Stream fairness, fairness between all streams in the system, is a more restrictive condition than \mbox{sub-stream} fairness,
fairness between all streams of each user. Thus \mbox{sub-stream} fairness alleviates utility loss as well as complexity and
overhead compared to stream fairness. Moreover, depending on algorithmic parameters, conventional algorithms including
distributed interference alignment (DIA) may not provide \mbox{sub-stream} fairness, and generate \mbox{sub-streams} with poor
signal-to-interference plus noise ratios (SINRs), thus with poor bit error rates (BERs). To this end, we propose a distributed
power control algorithm to render \mbox{sub-stream} fairness in the system, and establish initiatory connections between
\mbox{sub-stream} SINRs, BERs, and rates. Algorithms have particular responses to parameters. In the paper, important
algorithmic parameters are analyzed to exhibit numerical correctness in benchmarking. The distinction between separate
filtering schemes that design each stream of a user separately and group filtering schemes that jointly design the streams of
a user is also underscored in the paper. Finally, the power control law used in the proposed algorithm is proven to linearly
converge to a unique fixed-point, and the algorithm is shown to achieve feasible SINR targets.
\end{abstract}

\section{Introduction} \label{sec:Intro}
Fairness is important to ensure quality-of-service (QoS) in the system \cite{213,214}. While power control to attain
\mbox{signal-to-interference} plus noise ratio (SINR) fairness is well explored for downlink channels \cite{206,218}, the
research for interference channels (ICs) is still at a primitive level due to NP-hardness of the problem in general
\cite{211}. Fairness in the system can be achieved by two complementary approaches, maximization of minimum SINR subject to
power constraint or minimization of power subject to SINR constraint, and at three different levels, fairness between streams,
users, or \mbox{sub-streams}. Explicitly, fairness between all streams, all users, or all streams of each user can be aimed.
Both problems achieve optimal solutions when, depending on the intended level, streams', users', or \mbox{sub-streams'}
signal-to-interference plus noise ratios (SINRs) are attained with equality \cite{217}. From more to less restrictive, stream,
user, and \mbox{sub-stream} fairness come in order. Consequently, \mbox{sub-stream} fairness causes the least degradation in
\mbox{sum-rate}, followed by user and stream fairness. \mbox{Sum-rate} is a prominent but only a commensurate metric with each
stream's SINR. In particular, a stream can contribute substantially to \mbox{sum-rate}, but can have poor SINR, thus have
gross bit error rate (BER) \cite{220}. For example, SINR ratios of \mbox{sub-streams} can range from 5 to 9 times, whereas
rate ratios can be around 1.5 times only \cite{220}. As well known, although SINR and rate metrics are coupled via the log
function, the beamforming vectors that maximize the \mbox{sum-rate} do not necessarily maximize the \mbox{sum-SINR}. To this
end, we propose an \mbox{ad-hoc} distributed power control algorithm (DPCA) to achieve SINR fairness at the \mbox{sub-stream}
level by using the later approach, minimization of power subject to SINR constraint. Basically, transmit and receive
beamforming vectors are initially obtained via a conventional beamforming scheme including SINR maximization (\mbox{max-SINR})
and distributed interference alignment (DIA) \cite{55}. Then, our proposed power control algorithm is plugged and run in an
\mbox{ad-hoc} manner. The outer loop of the proposed algorithm linearly searches for a feasible SINR target for each user,
thus convergence is guaranteed. Since there is a maximum power constraint, the optimal power values may not be feasible if
SINRs are not well balanced before power control applied. The goal of the proposed algorithm is to achieve \mbox{sub-stream}
fairness while causing the least \mbox{sum-rate} degradation. Therefore, power saving is not the primary concern of our
algorithm. The power control law used in the algorithm to achieve \mbox{sub-stream} fairness in ICs is a direct extension of
standard power control law introduced in \cite{207}. However, our paper establishes initiatory connections between
\mbox{sub-stream} SINRs, bit error rate (BERs), and rates to some extent. In addition, our paper is the only paper except
\cite{223} to consider SINR targets as optimization variables as opposed to conventional approaches that propose schemes with
preset SINR or rate targets.

Achieving \mbox{sub-stream} fairness in physical layer, e.g., scheduling is not considered, is addressed in this paper.
Modifying conventional schemes including minimization of maximum mean square error \mbox{(min-max-MSE)} \cite{209} and
maximization of the proportional utility function \cite{266} so to achieve \mbox{sub-stream} fairness can be rewarding
compared with our simplistic proposed algorithm that suits practical applications. \mbox{Min-max-MSE} is computationally
costly, and maximization of proportional utility is NP-hard even for \mbox{multiple-input} \mbox{single-output} (MISO) ICs,
but in fact efficient algorithms can be proposed. However, in this paper, our goal is to take initiative steps on emphasizing
the importance of \mbox{sub-stream} fairness in the system and show its effects on system metrics, along with underscoring the
importance of numerical details. Thus this paper provides the basis for designing more powerful schemes to achieve
\mbox{sub-stream} fairness.

The prominent paper \cite{85} proposed a new technique that was coined interference alignment (IA), and IA was shown to
achieve the upper bound of non-interfering signaling dimensions in an interference channel (IC). Later, the limits of IA in
multiple-input multiple-output (MIMO) ICs were mainly identified in \cite{188} and in consecutive papers \cite{166,167}. The
numerical results of IA first appeared in \cite{55}, where the authors proposed a DIA algorithm. Subsequently modified DIA
techniques that achieved improved \mbox{sum-rate} performances appeared in \cite{158,219}. In \cite{55}, IC extension of the
conventional \mbox{max-SINR} algorithm for point-to-point channels was also proposed, and \mbox{max-SINR} was shown to achieve
higher \mbox{sum-rate} than DIA in the low to medium signal-to-noise ratio (SNR) regime. Although \mbox{max-SINR} lacks art in
its design approach, its \mbox{sum-rate} results in homogenous MIMO ICs (MIMO IC is fully connected  and channels are
independent and identically distributed, i.i.d.) is surprisingly satisfactory. Excluding the weighted minimum mean square
error (MMSE) technique \cite{187} proposed for single stream MIMO ICs, designing a linear scheme that superposes
\mbox{max-SINR} is still an open problem \cite{220}. However, an important QoS metric, \mbox{sub-stream} fairness, i.e.,
fairness between streams of a user, was overlooked in all these works. It can be shown that as the number of streams per user
increases, the SINR levels of some \mbox{sub-streams} can go very low implicating high decoding errors for those
\mbox{sub-streams}. This phenomena can occur due to the inherent competition in the algorithm (e.g., separate stream design of
\mbox{max-SINR}) and due to the preset parameters in the algorithm (e.g., stopping criterion of DIA). The results that showed
the importance of \mbox{sub-stream} fairness and algorithmic parameters first appeared in \cite{220}.

In the literature, \mbox{sub-stream} fairness was only studied in limited number for the layers above physical layer. Since
these papers are out of our scope, we do not cite them in the paper. To the best of our knowledge, our work is the first paper
to consider \mbox{sub-stream} fairness in physical layer. Our goal is to achieve fairness at \mbox{sub-stream} level in the
system with minimum \mbox{sum-rate} degradation. Algorithmic parameters shift the results considerably. For example, a
predetermined iteration number or a negligible increment in the \mbox{sum-rate} can be the stopping criterion of an algorithm.
While DIA can reasonably achieve \mbox{sub-stream} fairness for the later, the imbalance between \mbox{sub-streams} increases
as the preset iteration number decreases. In fact, for homogenous ICs, in ergodic sense, the three different fairness
approaches have similar outcomes but with complexity and overhead varieties. However, for a given channel, the three
approaches have disparate outcomes, especially in the high SNR regime and with a low preset iteration number \cite{220}. In
other words, for a given homogenous IC, altruism degree matters. For example, stream fairness, thus user fairness, can be
achieved at the cost of higher utility loss than \mbox{sub-stream} fairness. On the other hand, \mbox{sub-stream} fairness can
be achieved while letting users preserve their ranks, i.e., user fairness is not achieved, with the reward of lesser
complexity and overhead than stream fairness. Clearly, \mbox{sub-stream} fairness is less altruistic than stream fairness. For
example, for a given channel, assume the system has SINR distribution with
$\text{SINR}_\text{sys}=\{\{1,3\},\{4,6\},\{7,9\}\}$, where $\text{SINR}_k=\{\text{SINR}_{k,1},\text{SINR}_{k,2}\}$ denotes
\mbox{sub-stream} SINRs of user $k$. Please note that the average SINR per stream and user is 5 and 10, respectively. The
system can achieve $\text{SINR}_\text{sys-s}=\{\{3,3\},\{3,3\},\{3,3\}\}$,
$\text{SINR}_\text{sys-u}=\{\{3,5\},\{3,5\},\{3,5\}\}$ and $\text{SINR}_\text{sys-s-s}=\{\{2,2\},\{5,5\},\{7,7\}\}$ for
stream, user and \mbox{sub-stream} fairness, respectively.

Power control and beamforming are common approaches to achieve fairness. The reader is referred to \cite{204} for beamforming
design and to \cite{210} for joint power control and beamforming design, and the references therein. In \cite{204} and
\cite{210,211}, authors propose decentralized and centralized algorithms for ICs, respectively. In this work, we focus on
decentralized algorithms in consensus with ICs' nature. In downlink channels, SINR duality can be achieved between downlink
and uplink directions via various joint designs as summarized in \cite{221}. At each iteration, SINRs are equalized via power
control and incremented via beamforming, thus SINRs are maximized monotonically. Deploying this concept to ICs in a
distributed manner is nontrivial, thus designing joint power control and beamforming with decentralized and linear features to
achieve fairness in MIMO ICs is still an open problem. In this paper, we propose a practical and a distributed power control
algorithm that can be attached to any conventional beamforming scheme, i.e., the proposed algorithm is \mbox{ad-hoc}, with a
slightly increased algorithmic load. Simulation results show that the algorithm has narrow rate and SINR losses with achieved
\mbox{sub-stream} fairness. It is shown in \cite{220} that some \mbox{sub-streams} can have high decoding errors due to their
poor SINR levels, but their contribution to \mbox{sum-rate} can still be substantial. Therefore, \mbox{sum-rate} results
without parsed stream BERs cannot capture the whole picture. Our results also underline the ascending importance of
\mbox{sub-stream} fairness as \mbox{sub-stream} number increases.

Finally, a power control algorithm that swiftly converges to a \mbox{fixed-point} is requisite in practice. By using recently
introduced contractive interference functions \cite{225}, i.e., slightly modified versions of the well-known standard
interference functions \cite{207}, we prove that the power law in the proposed algorithm has a linear convergence rate to a
unique fixed-point making the algorithm preferable in practice.

The rest of the paper is organized as follows. In Section \ref{sec:Model}, we introduce the system model. In
Section \ref{sec:Fairness}, we focus on \mbox{sub-stream} fairness and give the motivation. In this section, we
present separate filtering schemes that design each stream of a user separately and group filtering schemes that
jointly design streams of a user, and introduce our power control algorithm. In Section \ref{sec:Parameters}, we
introduce important algorithmic parameters that can significantly differentiate numerical results and show that
a complete picture can only be depicted by varying these parameters in a benchmarking process. In Section
\ref{sec:Results}, we present the numerical results. In Section \ref{sec:ROC}, we show that the proposed
algorithm linearly converges to a unique fixed-point.

Notation: $^\dagger$ and $^{-1}$ denote the complex conjugate and inverse matrix (if the matrix is full-rank)
operations, respectively. Matrices are denoted by bold-face uppercase letters whereas vectors by bold-face
lowercase letters. $\mathbf{1}$, $\xI$, $\mathbf{0}$, and diag$[v_1,\ldots,v_l,\ldots, v_L]$ denotes all ones
vector, identity matrix,  zero vector or matrix, and diagonal matrix with elements $v_l$ on its diagonal,
respectively. $|.|$, $||.||_1$, and min denote determinant, $l_1$-norm, and minimum operators, respectively, and
for a given vector $\xv>\mathbf{0}$, $||.||_\infty^{\xtv}$ denotes weighted maximum norm.

\section{System Model} \label{sec:Model}

We consider a $K$-user IC, where there are $K$ transmitters and receivers with $M_k$ and $N_k$ antennas at node $k$,
respectively. A transmitter has $d_k$ streams to be sent to its corresponding receiver. This system can be modeled as
$\xy_k=\sum_{l=1}^K\xH_{kl}\xx_l+\xz_k, ~ \forall k\in\mathcal{K}\triangleq\{1,2,...,K\},$ where $\xy_k \textrm{ and } \xz_k$
are the $N_k\times 1$ received signal vector and the zero mean unit variance circularly symmetric additive white Gaussian
noise vector (AWGN) at the $k^{th}$ receiver, respectively. $\xx_l$ is the $M_l\times 1$ signal vector transmitted from the
$l^{th}$ transmitter and $\xH_{kl}$ is the $N_k\times M_l$ matrix of channel coefficients between the $l^{th}$ transmitter and
the $k^{th}$ receiver. $\xE[||\xx_l||^2]=p_l$ is the power of the $l^{th}$ transmitter. The transmitted signal from the
$l^{th}$ user is $\xx_l=\xU_l\sqrt{\xP_l}\xd_l$, where $\xU_l=[\xu_{l,1},\ldots,\xu_{l,d_l}]$ is the $M_l\times d_l$ precoding
(beamforming) filter, $\xd_l$ is $d_l\times 1$ vector denoting the $d_l$ independently encoded streams, and
$\xP_l=\text{diag}[p_{l,1},\ldots,p_{l,d_l}]$ is a $d_l\times d_l$ diagonal matrix consisting of \mbox{sub-stream} powers,
$p_l=\sum_{j=1}^{d_l}p_{l,j}$. The $N_l\times d_l$ receiver matrix is denoted by $\xV_l$.

\section{Sub-stream Fairness} \label{sec:Fairness}

In this section, we give the motivation for achieving \mbox{sub-stream} fairness in MIMO ICs, and then take preliminary steps
to introduce our algorithm. Particulary, we highlight the distinction between separate and group filtering schemes, and
introduce a slightly modified SINR definition for separate filtering schemes. Based on this new definition, we present the
standard interference function to be used in our proposed algorithm. Finally, we introduce our DPCA. We first begin with
summarizing an algorithm that also dynamically sets the SINR targets.

In the literature, SINR targets are generally predetermined \cite{211,222}, and to the best of our knowledge,
setting SINR targets opportunistically is only studied in \cite{223} by using the augmented Lagrangian penalty
function (ALPF) method. The method imposes fairness constraint between streams
\begin{equation*}
\xtSINR\xmkl-\text{min}~\textbf{SINR}^\ast_k\leq F,~\forall k \in \mathcal{K},~ \forall
l\in\mathcal{L}\triangleq\{1,2,...,d_k\},
\end{equation*}
where $\textbf{SINR}_k=\left[\xtSINR_{k,1},\ldots,\xtSINR_{k,d_k}\right]$ is the vector of \mbox{sub-stream} SINRs of user
$k$, $\textbf{SINR}_k^\ast$ is the vector excluding the compared SINR, i.e., $\textbf{SINR}_k^\ast=\textbf{SINR}_k\backslash\{
\xtSINR\xmkl\}$, and \textit{F} is the fairness constraint. \textit{F} can be called SINR fairness offset, and it clearly
limits the SINR difference between the streams. This method is a fair benchmark to our proposed algorithm since SINR targets
are set as optimization variables as well. However, the ALPF algorithm is not \mbox{ad-hoc}, thus a good starting point for
target SINR searching is critical for the algorithm. In \cite{223}, the authors use nonlinear search methods to find feasible
starting points. An important advantage of \mbox{ad-hoc} algorithms is their ability of searching target SINRs linearly.
Maximum \mbox{sub-stream} SINR per user achieved after beamforming is the upper bound to the maximum \mbox{sub-stream} SINR
achieved after power control. Therefore, setting average SINR per user as the \mbox{sub-stream} SINR target is a good starting
point for searching. ALPF algorithm can be extended to MIMO ICs and converted to an \mbox{ad-hoc} DPCA, but the extended
algorithm is expected to be considerably slower than our DPCA since they use Lagrangian function whereas we use a simple
standard interference function \cite{216} for power control as will be explained in the end of this section.

Before proceeding further, the reader is recommended to con our earlier paper \cite{220} that precedes the
current paper. Unbeknownst to us, a study that uses standard interference function was published \cite{224}
concurrently with the submission of our earlier work \cite{220}. The paper \cite{224} follows a conventional
motivation by proposing a power control algorithm to achieve predetermined rate targets per user with minimum
power consumption, and serves as a rectification to an erroneous SINR definition in \cite{212} as noted in
\cite{220}.

\subsection{Motivation}

\begin{figure}[htb]
\centering \subfigure[Iter=$\emptyset$] {
\includegraphics[height=5.25cm, width=9cm] {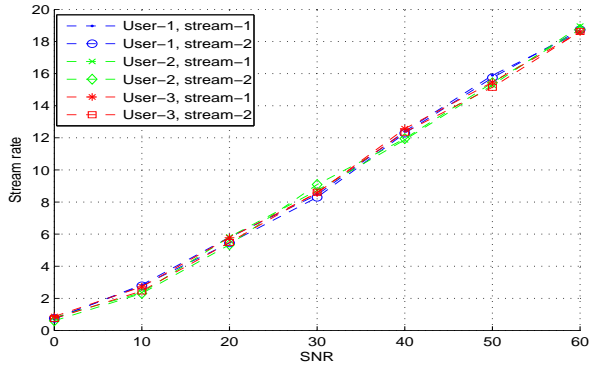} \label{Fig:Fig1a}
}
    \subfigure[Iter=50] {
\includegraphics[height=5.25cm, width=9cm] {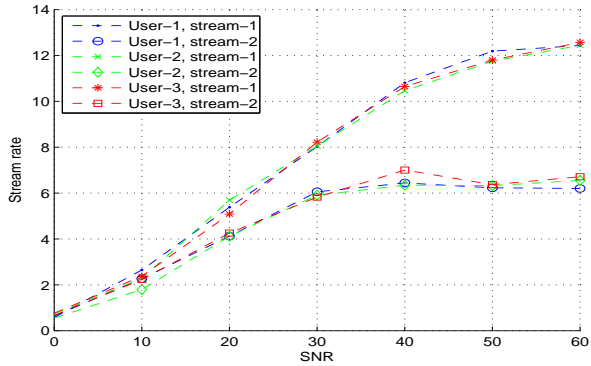} \label{Fig:Fig1b}
} \caption{Stream rates of DIA for different stopping criteria.} \label{Fig:Fig1}
\end{figure}

\begin{figure}[htb]
\centering \subfigure[Iter=$\emptyset$] {
\includegraphics[height=5.25cm, width=9cm] {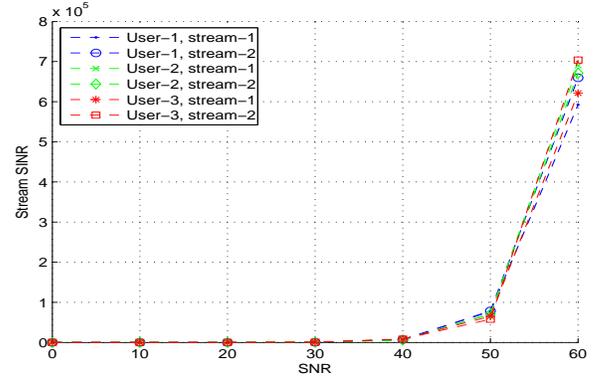} \label{Fig:Fig2a}
}
    \subfigure[Iter=50] {
\includegraphics[height=5.25cm, width=9cm] {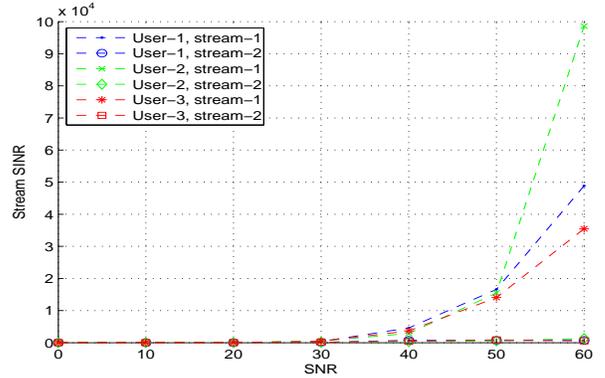} \label{Fig:Fig2b}
} \caption{Stream SINRs of DIA for different stopping criteria.} \label{Fig:Fig2}
\end{figure}

The importance of \mbox{sub-stream} fairness and algorithmic parameters is demonstrated in Fig. \ref{Fig:Fig1} and
\ref{Fig:Fig2} via Monte Carlo (MC) simulations using 40 independent channel realizations, MC=40. Unless otherwise stated in
the paper, we present numerical results for the IC with $K=3, M_k=4, N_k=4,\text{ and } d_k=2$, $(4\times4,2)^3$. The
abbreviation Iter in the plots stands for the number of iterations, basically a downlink and an uplink iteration are counted
as one. Iter=$\emptyset$ indicates that the number of iterations is not predetermined, and the algorithm stops when the
increment in the \mbox{sum-rate} is negligible
\begin{equation}\label{eqn:SumRateTarget}
|R_{\text{sum}}(n+1)-R_{\text{sum}}(n)|\leq \epsilon,
\end{equation}
where $n$ stands for the iteration number, and $\epsilon$ is $10^{-6}$ in our simulations. For all simulations
in the paper, we fix the number of random transmit beamforming initializations to one, presented numerical
results are per channel use, SINR results and SNR values are in linear and dB scale, respectively.

As explained in \cite{220} and as shown in Fig. \ref{Fig:Fig1} and \ref{Fig:Fig2}, DIA achieves reasonable \mbox{sub-stream}
fairness for Iter$=\emptyset$, but for fixed number of iterations, DIA cannot achieve balanced \mbox{sub-streams}. Numerical
results of \mbox{max-SINR} and another conventional scheme minimization of the sum of mean-square errors (min-sum-MSE) are
given in \cite{220}. In \cite{220}, it is shown that \mbox{max-SINR} cannot achieve \mbox{sub-stream} fairness even for
Iter=$\emptyset$, whereas \mbox{min-sum-MSE} still preserves fairness at a much better level even for a small number of
iteration, Iter=50. As known, BER is influenced by the worst stream SINRs in the system. Thus, in general, min-sum-MSE can
provide lower BER than \mbox{max-SINR} due to its stream fairness, and \mbox{max-SINR} can achieve lower BER than DIA since it
additionally aims to maximize desired signal power. In \cite{204}, BER performances are compared for a small iteration number,
Iter=16. Since the imbalance between streams in \mbox{max-SINR} and DIA soar marginally more than min-sum-MSE as the iteration
number decreases, BER gaps between these schemes are significant at Iter=16. These findings again indicate the influence of
iteration number, as other algorithmic details do, on perceiving the complete picture.

In \cite{220}, the results of \mbox{max-SINR} algorithm with orthonormal beamforming vectors are presented. In \cite{186,159},
it is observed that without the orthogonalization step, the algorithm yields linearly dependent beamforming vectors. In other
words, at least one stream of a user is shut off, thus a major \mbox{sum-rate} loss at high SNR is observed \cite{159}. Note
that MMSE receivers are assumed in \cite{186,159}, whereas we assume same type of filter structures are used at transmitters
and receivers. For either of these assumptions, we could not observe dependent beamforming vectors as in \cite{186,159}, but
the following. Algorithms require more iterations in the high SNR regime \cite{226,227}. However, \mbox{max-SINR} without the
orthogonalization step needs significantly high number of iterations, otherwise the algorithm saturates in the high SNR
regime. Please note that required iteration number and rate of convergence (RoC) of an algorithm are coupled parameters. More
on these parameters are discoursed in Sections \ref{sec:Parameters} and \ref{sec:ROC}.

In Table \ref{tabl:Table1}, the sum of SINR ratios of the $i^\text{th}$ to the $j^\text{th}$ stream,
$\sum_{k=1}^{K}(\text{SINR}_{k,i}/\text{SINR}_{k,j})$, of DIA ($i=1 \text{ and } j=2$), \mbox{max-SINR} ($i=2 \text{ and }
j=1$), and min-sum-MSE ($i=2 \text{ and } j=1$) for Iter=$50,\emptyset$, and 50, respectively, are given for MC=40. As seen
from the SINR and rate results of \mbox{max-SINR} in Table \ref{tabl:Table1} of this paper and in Fig. 1(a) of \cite{220},
some \mbox{sub-streams} can have low SINRs, thus high BERs, but their contribution in \mbox{sum-rate} can be substantial.
Please note that the first stream of DIA achieves higher SINR simply because beamforming vectors are assigned with
eigenvectors having in order from smaller to larger eigenvalues of interference covariance matrix \cite{55}. On the other
hand, the second stream of \mbox{max-SINR} achieves higher SINR because Gram-Schmidt approach is used for QR decomposition.

\begin{table} [htb!]
  \begin{center}
  \caption{Sum of SINR ratios.} \label{tabl:Table1}
\begin{tabular}{|c|c|c|c|c|c|c|c|c|}
  \hline
   \textit{\textbf{dB}} & \textbf{0} & \textbf{10} & \textbf{20} & \textbf{30} & \textbf{40}  & \textbf{50}  & \textbf{60} \\
     \hline
\textit{DIA} & 1.18 & 1.23 & 2.07 & 3.19 & 7.45 & 21.88 & 70.94 \\
     \hline
\textit{Max-SINR} & 1.74 & 2.92 & 4.12 & 5.87 & 6.17 & 7.21 & 7.35 \\
     \hline
\textit{Min-sum-MSE} & 1.03 & 1.04 & 0.94 & 1.02 & 0.79 & 0.76 & 1.40 \\
     \hline
\end{tabular}
    \end{center}
\end{table}
\subsection{Separate and Group Filtering Schemes} \label{sec:SFandGF}
This section starts by defining the SINR of a stream. Stream SINR can be defined in two ways based on whether separate
filtering (SF) or group filtering (GF) is applied \cite{220}. SF methods such as \mbox{max-SINR} design each stream of a user
separately, thus \mbox{sub-streams} are considered as interference on one another
\begin{equation}\label{eqn:SINRSF}
\xtSINR^\xSF\xmkl=\frac{\xv\xmkl^\dagger\xR\xmkl\xv\xmkl}{\xv\xmkl^\dagger\xB\xmkl\xv\xmkl},
\end{equation}
where $\xB\xmkl=\xQ\xmkl+\xI_{N_k}$,
$\xQ\xmkl=\sum_{j=1}^K\xH_{kj}\xU_j\xP_j\xU_j^\dagger\xH_{kj}^\dagger-\xR\xmkl$, and
$\xR\xmkl=p\xmkl\xH_{kk}\xu\xmkl\xu\xmkl^\dagger\xH_{kk}^\dagger$ are interference plus noise, interference
covariance, and covariance matrices of the $l^{th}$ stream of user $k$, respectively. Note that $\xB\xmkl$
contains the \mbox{intra-user} interference. \mbox{Max-SINR} is apparently \mbox{sub-optimal} even for a
\mbox{point-to-point} system with multiple streams since the streams are competing with each other. A better
approach is to design beamforming vectors by allowing cooperation between them \cite{176}. Group filtering
methods such as generalized eigenvalue decomposition (GEVD) and semidefinite programming (SDP) \cite{197}
jointly design the streams of a user, thus allowing collaboration between beamforming vectors of a user
\cite{220}
\begin{equation}\label{eqn:SINRGF}
\xtSINR^{\xGF}\xmkl=\frac{\xv\xmkl^\dagger\xR_k\xv\xmkl}{\xv\xmkl^\dagger\xB_k\xv\xmkl},
\end{equation}
where $\xB_k=\xQ_k+\xI_{N_k}$, ${\xQ_k=\sum_{j=1,j\neq k}^K\xH_{kj}\xU_j\xP_j\xU_j^\dagger\xH_{kj}^\dagger}$, and
$\xR_k=\xH_{kk}\xU_k\xP_k\xU_k^\dagger\xH_{kk}^\dagger$ are interference plus noise, interference covariance, and covariance
matrices of user $k$, respectively. For the \mbox{max-SINR} algorithm, it is shown in \cite{220} that there is nearly no
difference between reckoning and not reckoning \mbox{intra-user} interference when Shannon's rate formula is used. In fact,
for ICs, joint decoders such as maximum likelihood (ML) decoder are preferred since linear decoders such as hard decision
decoder achieves severely poor BERs. Therefore, the motivation of \mbox{max-SINR} to consider \mbox{intra-user} interference
is not clear. Consequently, for ICs, we use the following beamforming filter for the \mbox{max-SINR} algorithm
\begin{equation} \label{eqn:newmaxsinrfilter}
\xv\xmkl=\frac{\xB^{-1}_{k}\xH_{kk}\xu\xmkl}{||\xB^{-1}_{k}\xH_{kk}\xu\xmkl||}.
\end{equation}
In line with the above beamforming filter, the new SINR definition for SF schemes used in ICs can be given as
\begin{equation}\label{eqn:SINRSFprime}
\xtSINR^{\xSF'}\xmkl=\frac{\xv\xmkl^\dagger\xR\xmkl\xv\xmkl}{\xv\xmkl^\dagger\xB_{k}\xv\xmkl}.
\end{equation}
The justification of \eqref{eqn:newmaxsinrfilter} and \eqref{eqn:SINRSFprime} can also be shown in a different approach as
follows. Consider the conventional \mbox{max-SINR} filter that reckons \mbox{intra-user} interference, thus the SINR
definition in \eqref{eqn:SINRSF} is used. Now consider a GF scheme, thus SINR definition in \eqref{eqn:SINRGF} is used. As
explained in \cite{220}, SF and GF achieve similar rate results when Shannon's formula is used. On the other hand, the SINRs
that SF and GF schemes achieve significantly deviate due to the difference in SINR formulas \eqref{eqn:SINRSF} and
\eqref{eqn:SINRGF} while numerical results show that these schemes achieve similar BERs. These results indicate that
evaluating beamforming filters and SINRs of SF schemes with ML decoders by \eqref{eqn:newmaxsinrfilter} and
\eqref{eqn:SINRSFprime} is more congenial. Note that when compared to $\xR\xmkl$ \eqref{eqn:SINRSFprime}, GF schemes can
explore the extra degrees of freedom in $\xR_k$ \eqref{eqn:SINRGF}.

\subsection{Standard Interference Function} \label{sec:SIF}
We briefly recall the well-known DPCA with maximum power per user $p_k$ constraint \cite{213}
\begin{equation}\label{eqn:minop}
p_k^n=\text{min}\left(\frac{\Gamma_k}{\xtSINR_k^{n-1}}p_k^{n-1},p_k\right),
\end{equation}
where superscript $n$ is the iteration number, $p_k^n$ is the power, $\xtSINR_k^{n-1}$ is the SINR of user $k$,
and $\Gamma_k$ is the SINR target. As seen in \eqref{eqn:minop}, each user updates its power following a simple
decision function, i.e., min operator. Basically, a user increases its power if its SINR is below its SINR
target and vice versa. Clearly the SINR target can be unmet due to the maximum power constraint.

Using the SINR constraint per stream $\xtSINR^{\xSF'}\xmkl\geq\Gamma_k$, the interference function for our
problem is given as
\begin{equation}\label{eqn:ourSIF}
\xtI\xmkl(\xp)=\Gamma_k\delta\xmkl,
\end{equation}
where $\xp=[p_{1,1},\ldots,p_{1,d_1},\ldots,p_{K,1},\ldots,p_{K,d_K}]$ is the power vector of the system,
$\Gamma_k=\overline{\text{SINR}}_k$ is the SINR target set to average SINR of user $k$, and
\begin{equation}\label{eqn:newterm}
\delta\xmkl=\frac{p\xmkl}{\xtSINR^{\xSF'}\xmkl}.
\end{equation}
The interference function $I\xmkl(\xp)$ can be shown to be standard, i.e., interference function satisfies monotonicity and
scalability properties, following a similar approach in \cite{224}. As mentioned in Section \ref{sec:Intro}, satisfying SINR
duality, i.e., carrying a \mbox{two-way} joint design approach, is nontrivial in ICs. Presumably owing to this fact, the
proposed algorithm in \cite{224} is also a \mbox{one-way} but joint power control and beamforming design, viz. power control
is applied in only downlink direction after each beamforming design step. Finally, the new SINR definition introduced in this
section will play an important role in Section \ref{sec:ROC} for proving the RoC of the proposed algorithm.

\subsection{Proposed Algorithm}
Before presenting our algorithm, we first review some possible approaches to achieve \mbox{sub-stream} fairness.
For joint power control and beamforming design to achieve \mbox{sub-stream} fairness, the problem can be
formulated as
\begin{align}\label{eqn:P1MaxMin}
    &\max_{\xU_k,\xV_k,\xp_k}\min_l\frac{\xtSINR^{\xSF'}\xmkl}{\Gamma_k}
    \\ &\text{~subject to~} \sum_{l=1}^{d_l}p\xmkl\leq p_k,
    \forall k\in\mathcal{K} \text{ and } \forall l\in\mathcal{L}. \nonumber
\end{align}
Compared with stream fairness problems, the \mbox{sub-stream} problems are decoupled into $K$ \mbox{sub-problems} given SINR
targets are feasible, thus the above problem is solved asynchronously among users. However, feasibility check is coupled among
users and can be shown to be \mbox{NP-hard} \cite{211}. Therefore, we focus on designing efficient algorithms for achieving
locally optimal points. Please note that the cyclic coordinate ascent algorithm presented in \cite{211} can be modified to
achieve \mbox{sub-stream} fairness. This method is a fair benchmark to our proposed algorithm since it can also be applied in
a distributed manner, but its complexity is higher than our simplistic algorithm. A centralized approach can be maximization
of sum of all stream SINRs while SINR target per \mbox{sub-stream} and power constraint per user are met. Developing more
advanced schemes for \mbox{sub-stream} fairness is our next research direction.

As well known, and mentioned before, locally optimal solutions are obtained when SINR constraints are active, in other words
when SINR constraints are satisfied with equality. Moreover, without the min operator, the problem \eqref{eqn:P1MaxMin} is
convex over transmit or receive filters and a closed-form solution exists \cite{55}, but the problem is not jointly convex
over all beamforming matrices. Motivated by these results, the problem \eqref{eqn:P1MaxMin} can be divided into two
sub-problems. Beamforming vectors can be first obtained via conventional schemes including \mbox{max-SINR} and DIA \cite{55}.
Then in the second sub-problem where beamforming vectors are fixed, and by applying active SINR constraints, power vectors can
be obtained
\begin{align}\label{eqn:P2MaxP}
&\max_{\xp_k} \Gamma_k\\ &\text{ subject to } \xtSINR^{\xSF'}\xmkl=\Gamma_k, \sum_{l=1}^{d_l}p\xmkl\leq p_k,
\forall k\in\mathcal{K} \text{ and } \forall l\in\mathcal{L}. \nonumber
\end{align}
Given feasible SINR targets, the above problem \eqref{eqn:P2MaxP} is still nontrivial \cite{210}. After shedding light on
current state of the problem, and addressing open points, we propose a practical scheme, named \mbox{ad-hoc} DPCA, to achieve
\mbox{sub-stream} fairness. Basically, we unite the simple linear search for finding maximum possible SINR targets with the
minimization of power subject to SINR constraint problem
\begin{equation*}
    \min_{\xp_k}p_k \text{ subject to } \xtSINR^{\xSF'}\xmkl\geq \Gamma_k, \forall k\in\mathcal{K} \text{ and } \forall l\in\mathcal{L}
\end{equation*}
that can be solved via conventional DPCA.

The proposed \mbox{ad-hoc} DPCA in Algorithm \ref{alg:AdHocAlg} opportunistically searches a feasible SINR
target for each user. The algorithm can run asynchronously among users. In Algorithm \ref{alg:AdHocAlg},
$\xp_k^n=[p_{k,1}^n,\ldots,p_{k,d_k}^n]$ is the power vector and $p\xmkl^n$ is the power for the $l^{th}$ stream
of user $k$ at iteration $n$, $p_k^n=\sum_{l=1}^{d_k}p\xmkl^n$. $\xB_k^{n-1}=\xQ_k^{n-1}+\xI_{N_k}$ and
$\xQ_k^{n-1}=\sum_{j=1,j\neq k}^K\xH_{kj}\xU_j\xP_j^{n-1}\xU_j^\dagger\xH_{kj}^\dagger$ are interference plus
noise and interference covariance matrices, respectively,
$\xR\xmkl^\prime=\xH_{kk}\xu\xmkl\xu\xmkl^\dagger\xH_{kk}^\dagger$ is akin to a covariance matrix,
$\xP_j^{n-1}=\text{diag}[p_{j,1}^{n-1},\ldots,p_{j,d_j}^{n-1}]$ is a diagonal matrix of \mbox{sub-stream}
powers, $\mathbf{1}=[1,\ldots,1]$ is all ones vector, and
$\boldsymbol{\delta}_k=[\delta_{k,1},\ldots,\delta_{k,d_k}]$ is the vector of terms defined in
\eqref{eqn:newterm}, respectively.

\begin{algorithm}[htb!]
\footnotesize{\caption{Ad-Hoc DPCA} \label{alg:AdHocAlg}
\begin{algorithmic}[1]
\State Evaluate SINR outcomes of a beamforming scheme, $\text{SINR}\xmkl$ \State initialize
$\text{SINR}\xmkl^\prime=\text{SINR}\xmkl$, $\forall k\in \mathcal{K}, \forall l\in\mathcal{L}
$\label{algsteps:InitSINRs} \State check=0 \While {check$\sim$=1} \State $\xp_k^0=\frac{p_k}{d_k}\mathbf{1}$,
$\xp_k^1=2\xp_k^0$, $\Gamma_k=\overline{\text{SINR}}_k^\prime$, $\forall k\in \mathcal{K}$
\label{algsteps:Initp} \State $n=1$ \While{$\sum_{k=1}^K||\xp_k^n-\xp_k^{n-1}||_1>\epsilon$} \State
$\delta\xmkl=\frac{\xv\xmkl^\dagger\xB_k^{n-1}\xv\xmkl}{\xv\xmkl^\dagger\xR\xmkl^\prime\xv\xmkl}$, $\forall k\in
\mathcal{K}, \forall l\in\mathcal{L} $ \State $x=2\max(\boldsymbol{\delta}_k)$, $p_k^T=0$, $\forall k\in
\mathcal{K}$ \For {counter=1:$d_k$}, $\forall k\in \mathcal{K}$ \State $[\sim,y]=\min(\boldsymbol{\delta}_k)$
\label{step:min} \State $p_{k,y}^n=\min(\Gamma_k\delta_{k,y},p_k-p_k^T)$ \State $p_k^T=p_k^T+p_{k,y}^n$,
$\delta_{k,y}=x$ \EndFor \State $n=n+1$\EndWhile \State Evaluate new SINRs $\text{SINR}\xmkl^\prime$ by using
new power values $\xp_k^n$, $\forall k\in \mathcal{K}$ \If
{$\sum_{k=1}^K\sum_{\substack{m,n=1\\
m\neq n}}^{d_k}|\text{SINR}_{k,m}^\prime-\text{SINR}_{k,n}^\prime|\leq\epsilon$} \State check=1 \EndIf \EndWhile
\end{algorithmic}}
\end{algorithm}

The outer while loop in Algorithm \ref{alg:AdHocAlg} searches a feasible SINR target for each user. Since there is a maximum
power constraint, the optimal power values may not be feasible if SINRs are not well balanced before power control applied.
For example consider the system $\text{SINR}_\text{sys}=\{\{2,6\},\{2,4\},\{1,15\}\}$. Clearly, the \mbox{sub-stream} SINRs of
the last user $k=3$  are not well balanced. The optimal power values may not be feasible to achieve fairness between
\mbox{sub-streams}, i.e., $\text{SINR}_\text{sys,sub-str,opt}=\{\{4,4\},\{3,3\},\{8,8\}\}$ may not be achieved, but
$\text{SINR}_\text{sys,sub-str}=\{\{4,4\},\{3,3\},\{5,5\}\}$ may be achieved instead. The step \ref{step:min} of the algorithm
is the most critical part where powers of \mbox{sub-streams} are updated in order from the \mbox{sub-stream} with the lowest
to the highest $\delta\xmkl$. This way the \mbox{sub-stream} with the lowest $\delta\xmkl$ can definitely reach the SINR
target, while the \mbox{sub-stream} with the highest $\delta\xmkl$ reaches to a maximum possible SINR value with the remaining
power budget of the user. In the next iteration, the target SINR is the average of these achieved SINRs, thus the algorithm
keeps iterating until the convergence of \mbox{sub-stream} SINRs. The convergence plot of the proposed algorithm is given in
\cite{220}. As shown in Section \ref{sec:ROC}, our proposed algorithm has a fast convergence rate, and each iteration has
small costs \cite{220}.

\section{Algorithmic Parameters} \label{sec:Parameters}

In this section, we briefly introduce some important algorithmic parameters that significantly affect the results and our
perceptions in benchmarking. Different algorithms have different responses to algorithmic parameters. Therefore carefully
scanning these parameters is important to benchmark entirely. For example, the \mbox{sum-rate} gap between \mbox{max-SINR} and
DIA in low to medium SNR regime can be emphasized or both can be asserted as not achieving \mbox{sub-stream} fairness when
screening of parameters shortfall. As shown in the previous section, iteration number is the critical factor for the later.
For the former, the number of initializations of random transmit beamforming vectors is the critical parameter. Numerical
results show that if more initializations are allowed, the \mbox{sum-rate} gap in the low to medium SNR regime between DIA and
\mbox{max-SINR} is reduced. This can indicate that local \mbox{sum-rate} optimal points of DIA are more inhomogeneously
distributed than those of \mbox{max-SINR}, thus more initializations can increase the chances of finding a better local
optimum for DIA.

The number of iterations required for an algorithm to converge is coupled with the algorithm's RoC as mentioned in Section
\ref{sec:Fairness}. RoC is shown to depend on SNR in \cite{226,227}, and it also depends on whether the algorithm provides
fair \mbox{sub-streams} or not. If the algorithm cannot provide fair \mbox{sub-streams}, an utmost example can be some streams
are completely shut off, as seen in DIA and \mbox{max-SINR} examples in previous sections, the algorithm can require less
number of iterations in general. Conversely, the algorithm requires more iterations, i.e., the algorithm's RoC is slower, when
for example the system size increases, e.g., the numbers of users and fair \mbox{sub-streams} are increased. \mbox{Sum-rate}
results for different $\epsilon$ values of the stopping criterion \eqref{eqn:SumRateTarget} are given in Table \ref{tabl:RoC}
for MC=20. For each scheme, the results in the first and second rows are for $\epsilon=10^{-6}$ and $10^{-2}$, respectively.
In the iteration number column of the table, the iteration number of the last MC trial for SNR 40 dB is given since in general
high SNR regime requires more iterations. In the (normalized) convergence speed column of the table, the \mbox{sum-rate}
difference normalized by the iteration number difference, e.g., $\frac{72.2-59.6}{358-147}$, is given. From slower to faster
convergence speed, \mbox{min-sum-MSE}, \mbox{max-SINR} without, \mbox{max-SINR} with orthogonalization step, and DIA come in
order. Basically, schemes that provide \mbox{sub-stream} fairness tend to be slower as in the cases of \mbox{min-sum-MSE} and
\mbox{max-SINR} without the orthogonalization step. The convergence speed of DIA is still high, although for Iter=$\emptyset$
it achieves quite fair \mbox{sub-streams} as mentioned before. \mbox{Max-SINR} is not a true SINR maximizer as shown in the
next section, thus \mbox{max-SINR} and \mbox{min-sum-MSE} can have slow convergence speeds for \mbox{sum-rate} maximization
objective. Whereas DIA aims for the minimization of interference, a more influential objective at high SNR, thus it can have a
fast convergence speed for \mbox{sum-rate} maximization.

\begin{table}
  \begin{center}
  \caption{Results for different $\epsilon$ values.} \label{tabl:RoC}
  \renewcommand{\arraystretch}{1.2}
\begin{tabular}[htb]{|c|c|c|c|c|c|c|c|}
  \hline
\textit{\textbf{dB}} & \textbf{0} & \textbf{10} & \textbf{20} & \textbf{30} & \textbf{40}  & \textbf{Iter.}  & \textbf{Conv.}
 \\
     \hline
 \multirow{2}{*}{\textit{DIA}} & 4.3 &   15.4 &  32.2 &  52.1  & 72.2 & 358 & 0.0602\\ \cline{2-8}
     & 3.7 &   13.6 &  27.1 &  38.2  & 59.6 & 147 & \\
 \hline
\multirow{2}{*}{\textit{Max-SINR w/}} & 9.3 &   21.8 &  37 &  53.3  & 75.4 & 340 & 0.0624\\ \cline{2-8}
     & 9.3 & 21.2 &  34.6 &  45.9  & 63.3 & 146 & \\
 \hline
\multirow{2}{*}{\textit{Max-SINR w/o}} & 9.4 & 22 &  37.2 &  51.5  & 71.6 & 12880 & 0.0015\\ \cline{2-8}
     & 9.3 & 21.5 &  34.7 &  44.6  & 52.4 & 50 & \\
 \hline
\multirow{2}{*}{\textit{Min-sum-MSE}} & 9.3 & 21.4 &  36.2 &  51.2  & 64.2 & 21153 & 0.0006\\ \cline{2-8}
     & 9.2 & 20.9 &  32.8 &  43.5  & 50.7 & 82 & \\
 \hline
\end{tabular}
    \end{center}
\end{table}

\mbox{Max-SINR} without the orthogonalization step can achieve reasonable \mbox{sub-stream} fairness in low to medium SNR
regime. However, at high SNR along with low preset iteration number, fairness cannot be achieved due to the inherent
competition between \mbox{sub-streams}. On the other hand, with the orthogonalization step, \mbox{max-SINR} cannot achieve
\mbox{sub-stream} fairness as shown in \cite{220} and in Section \ref{sec:Results} of this paper. Some streams achieve poor
SINR levels although their contribution in \mbox{sum-rate} is significant. Therefore, without observing streams with low SINRs
thus with high BERs, the poor \mbox{sum-rate} performance of \mbox{max-SINR} in the high SNR regime seems to be improved by
the orthogonalization step.

\begin{figure}[htb]
\centering
\includegraphics[height=5.25cm, width=9cm] {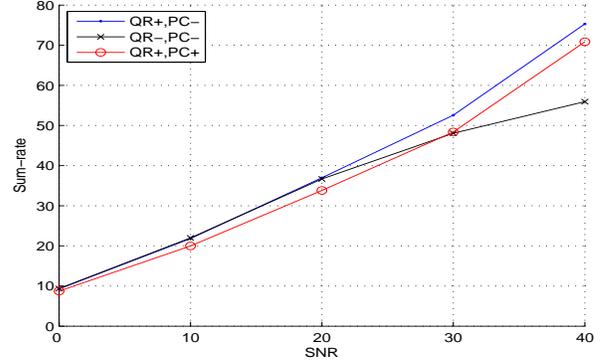}
 \caption{Sum-rates of max-SINR for different cases.} \label{Fig:SumRate}
\end{figure}

In Fig. \ref{Fig:SumRate}, the \mbox{sum-rate} results of \mbox{max-SINR} algorithm for different cases are presented for
MC=20 and Iter=1000. In the legend, we use + and - to indicate whether the feature exists or not, respectively. For example,
the (QR+,PC-) legend denotes the \mbox{max-SINR} scheme with orthonormal beamforming vectors but without the power control
(PC) algorithm. A simple QR decomposition can be used to obtain orthogonal beamforming vectors. As seen in the figure,
orthogonalization step seems to fix the high SNR region problem of conventional \mbox{max-SINR} algorithm that has no power
control. However, as will be shown in Section \ref{sec:Results}, \mbox{max-SINR} without power control generates
\mbox{sub-streams} with low SINRs, thus with high BERs. In the following sections, we use \mbox{max-SINR} algorithm with
orthonormal beamforming vectors since the RoC of \mbox{max-SINR} without orthogonalization step is much slower.

Finally, every detail in numerical results can be considered as algorithmic parameter, ranging from how \mbox{sum-rate} is
calculated, e.g., Shannon's or sum of stream rates formula is used, to MC number. As well-known, MC number, i.e., the total
number of tested channels that are independently generated, is another founding parameter to exhibit numerical correctness. A
low set MC number for plotting ergodic \mbox{sum-rate} and \mbox{sum-SINR} can give a false impression especially for
\mbox{sub-streams} as seen in Fig. \ref{Fig:Fig4} of this paper and in Fig. 4 of \cite{220}. While running BER simulations
that require high MC tests, we evaluate the SINR values as well, and present the results in Section \ref{sec:Results}.
Ultimately the ergodic rate and SINR results are found to be more accurate than those simulations with low MC number.

\section{Numerical Results} \label{sec:Results}

For all algorithms except min-sum-MSE \cite{204,177}, we assume the same type of filter structures are used at the
transmitters and receivers, e.g., \mbox{max-SINR} filter \eqref{eqn:newmaxsinrfilter} is used for both transmit and receive
filters. In \cite{204,177}, authors show that MMSE receive filter is optimal for the given transmit filter, and obtain optimal
transmit filter via Lagrangian solution. On the other hand, our approach is useful, for example, for sensor network
transceivers. In addition, our approach isolates the problem on the considered filter, e.g., \mbox{max-SINR} filter. For
better scrutinizing the rate and SINR results of the algorithms, we save randomly initiated beamforming vectors and channels
into files, and feed the same files into benchmarked algorithms. Since BER results require abundant number of random channels
to be tested, which increases the file sizes thus run time, we do not follow the same approach for BER results.

\subsection{Targeting Sum-SINR}
\mbox{Max-SINR} results that are presented in the literature are generally obtained via targeting \mbox{sum-rate}
maximization, e.g., algorithms stop based on condition \eqref{eqn:SumRateTarget}, as opposed to a possible perception that the
objective is \mbox{sum-SINR} maximization. Since there is no \mbox{one-to-one} correspondence between \mbox{sum-SINR} and
\mbox{sum-rate} metrics, beamforming optimization based on \mbox{sum-SINR} maximization yields lower \mbox{sum-rate}, and vice
versa. As mentioned before, improving the worst stream SINR in the system improves BER of the system. Therefore, the stopping
criterion
\begin{equation}\label{eqn:SumSINRTarget}
|\text{SINR}_{\text{sum}}(n+1)-\text{SINR}_{\text{sum}}(n)|\leq \epsilon,
\end{equation}
where $\text{SINR}_{\text{sum}}=\sum_{k=1}^K\sum_{l=1}^{d_k}\text{SINR}_{k,l}$ is the sum of SINRs, is also plausible. In
Table \ref{tabl:Table2}, \mbox{sum-SINR} results are presented. The starred and unstarred algorithm denotes the algorithm with
\mbox{sum-SINR} and \mbox{sum-rate} maximization objective, in other words with the stopping criterion
\eqref{eqn:SumSINRTarget} and \eqref{eqn:SumRateTarget}, respectively. The results are normalized with the unstarred GEVD
results. Although the starred \mbox{max-SINR} algorithm ($\text{max-SINR}^\ast$) truly aims for SINR maximization, starred DIA
($\text{DIA}^\ast$) achieves higher \mbox{sum-SINR}. Interestingly, we observe that the received signal power of
$\text{DIA}^\ast$ is more than $\text{max-SINR}^\ast$ as opposed to expected. On the other hand, interference signal power of
$\text{DIA}^\ast$ is slightly lesser than $\text{max-SINR}^\ast$. This is another indicator that \mbox{max-SINR} scheme is far
from being an optimal SINR maximizer.

\begin{table}
  \begin{center}
  \caption{Sum-SINRs normalized by GEVD results.} \label{tabl:Table2}
\begin{tabular}[htb]{|c|c|c|c|c|c|c|c|c|}
  \hline
   \textit{\textbf{dB}} & \textbf{0} & \textbf{10} & \textbf{20} & \textbf{30} & \textbf{40}  & \textbf{50}  & \textbf{60} \\
     \hline
  \textit{DIA$^\ast$} & 0.74 & 1.43 & 1.68 & 1.84 & 1.82 & 1.88 & 1.91 \\
     \hline
  \textit{DIA} & 0.65 & 0.87 & 0.97 & 1.00 & 0.99 & 0.97 & 0.99 \\
     \hline
  \textit{Max-SINR$^\ast$} & 1.01 & 1.05 & 1.07 & 1.05 & 1.05 & 1.10 & 1.08 \\
     \hline
  \textit{Max-SINR} & 0.99 & 1.00 & 1.01 & 1.00 & 1.00 & 0.98 & 1.01 \\
     \hline
  \textit{GEVD$^\ast$} & 1.01 & 1.04 & 1.04 & 1.05 & 1.05 & 1.10 & 1.06 \\
  \hline
\end{tabular}
    \end{center}
\end{table}


\subsection{BER Results}

\begin{figure}[htb]
\centering \subfigure[with power control] {
\includegraphics[height=4.5cm, width=8cm] {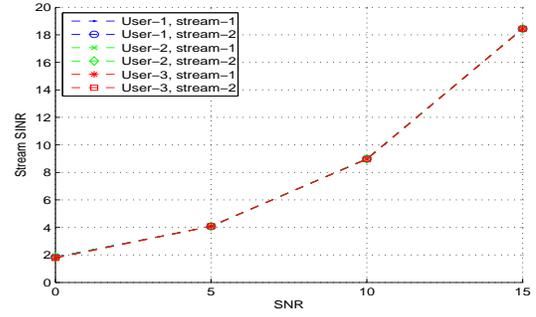} \label{Fig:Fig3a}
}
    \subfigure[without power control] {
\includegraphics[height=4.5cm, width=8cm] {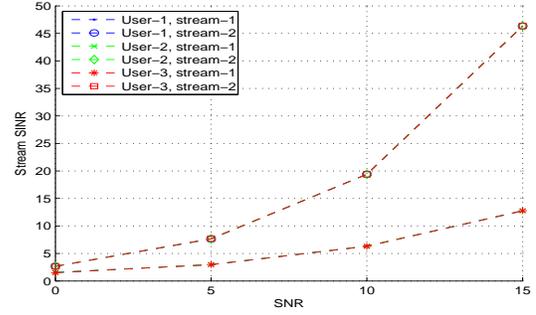} \label{Fig:Fig3b}
} \caption{Stream SINRs of max-SINR.} \label{Fig:Fig3}
\end{figure}

\begin{figure}[htb]
\centering
\includegraphics[height=4.5cm, width=8cm] {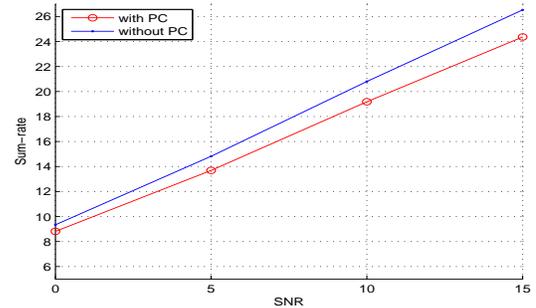}
 \caption{Sum-rates of max-SINR.} \label{Fig:Fig6}
\end{figure}

\begin{figure}[htb]
\centering \subfigure[with power control] {
\includegraphics[height=5.25cm, width=9cm] {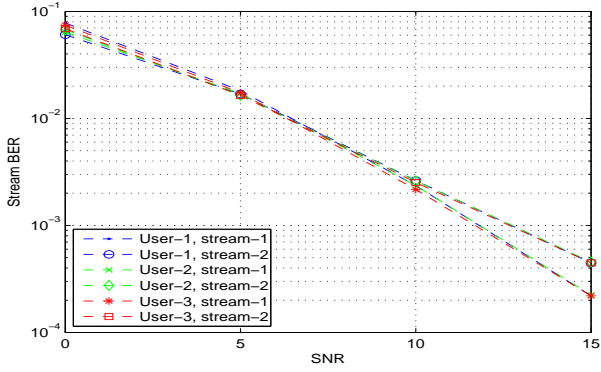} \label{Fig:Fig4a}
}
    \subfigure[without power control] {
\includegraphics[height=5.25cm, width=9cm] {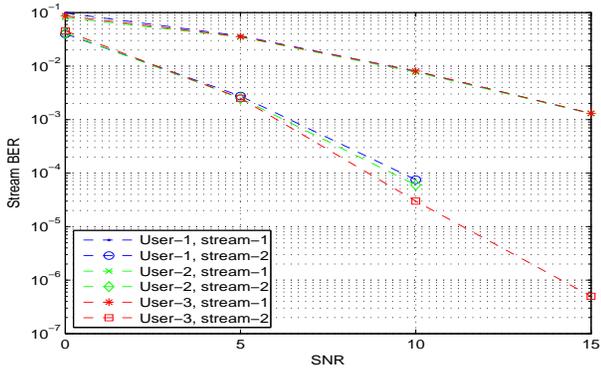} \label{Fig:Fig4b}
}
   \subfigure[BER comparison] {
\includegraphics[height=5.25cm, width=9cm] {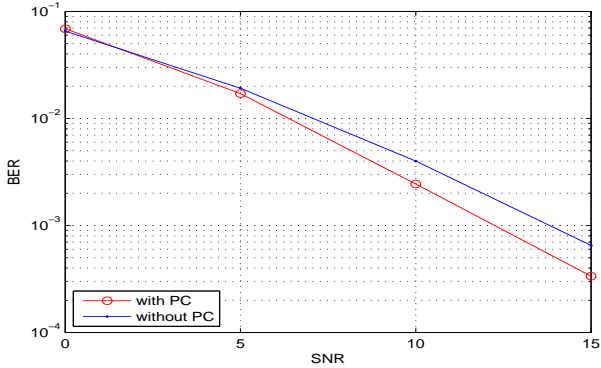} \label{Fig:Fig4c}
} \caption{Error rates of max-SINR. } \label{Fig:Fig4}
\end{figure}

Bit error, \mbox{sum-rate} and SINR results of \mbox{max-SINR} with and without the proposed power control in Algorithm
\ref{alg:AdHocAlg} for the system \mbox{$(4\times4,2)^3$} are presented in Fig. \ref{Fig:Fig3} - \ref{Fig:Fig4}, and BER
result for ${(6\times6,3)^3}$ is presented in Fig. \ref{Fig:Fig7}. To plot BER results in the paper, MC= $10^3, 10^4, 10^5,
\text{ and } 10^6$ random ICs are tested for SNR values 0, 5, 10, and 15 dB, respectively. Channel coefficients are generated
by i.i.d. zero-mean \mbox{unit-variance} complex Gaussian variables, QPSK modulation is used, and Iter=16 is chosen. As seen
in Fig. \ref{Fig:Fig3a}, the proposed \mbox{ad-hoc} DPCA whose objective is \mbox{sub-stream} fairness can achieve stream
fairness in ergodic sense with lesser algorithmic complexity and information exchange than the algorithms whose objectives are
stream fairness. As seen in Fig. \ref{Fig:Fig6}, \mbox{sub-stream} fairness is achieved at the cost of a reasonable
\mbox{sum-rate} degradation.

BER results of \mbox{max-SINR} with and without power control are given in Fig. \ref{Fig:Fig4}. Clearly, error rates in Fig.
\ref{Fig:Fig4} are coherent with SINR results in Fig. \ref{Fig:Fig3}. As seen in Fig. \ref{Fig:Fig4b}, for the given number of
trials, two streams have no bit errors at SNR 15 dB, thus not plotted. To achieve leveled BERs in Fig. \ref{Fig:Fig4a} similar
to leveled SINRs in Fig. \ref{Fig:Fig3a}, many more MC simulations needed even for SNR 0 dB. Due to lengthy simulation times,
we avoid such high MC numbers.

\begin{figure}[htb]
\centering
\includegraphics[height=4.5cm, width=8cm] {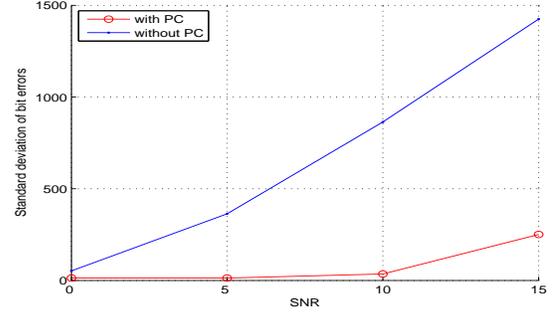}
 \caption{Standard deviations of streams' bit errors.} \label{Fig:Fig5}
\end{figure}

In Fig. \ref{Fig:Fig5}, standard deviation of bit errors per stream is given. Basically, total number of bit errors per stream
is obtained, and then standard deviation of these numbers is evaluated. Since more channels are tested as SNR increases, total
number of errors per stream, thus standard deviation increases. However, Fig. \ref{Fig:Fig5} clearly shows the success of DPCA
on keeping fairness in the system. In Table \ref{tabl:Table3}, total and average number of bit errors are explicitly given at
\mbox{sub-stream} and system levels. In the table, $\text{BE}_k=\{\text{BE}_{k,1},\text{BE}_{k,2}\}$ denotes the total number
of bit errors per stream of user $k$, e-$x$ denotes $10^{-x}$, which is a multiplicative factor, e.g.,
$\{8.035,7.5\}(\text{e-3,e-5})\triangleq\{8.035\times10^{-3},7.5\times10^{-5}\}$. From these results, we see that DPCA
enforces bit errors to be distributed more homogenously among the streams.

As the number of \mbox{sub-streams} increased, the imbalance between them increases, thus power control to achieve
\mbox{sub-stream} fairness becomes more imperative as seen in Fig. \ref{Fig:Fig7}.

\begin{figure}[htb]
\centering
\includegraphics[height=5.25cm, width=9cm] {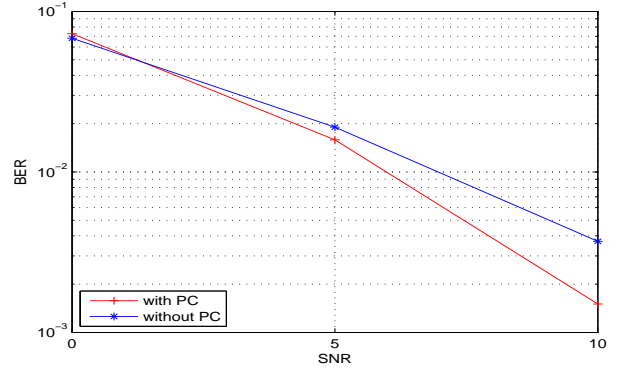}
 \caption{BERs of max-SINR for the $(6\times6,3)^3$ system.} \label{Fig:Fig7}
\end{figure}

\begin{table}[htb]
\scriptsize{
  \begin{center}
    \caption{Total and average number of bit errors.} \label{tabl:Table3}
\begin{tabular}{|p{.5cm}|p{.2cm}|l|c|c|}
 \hline
 \multicolumn{3}{|c|}{\textit{\textbf{Level}}} & \textbf{Sub-stream} & \textbf{System}\\ \hline

\multirow{8}{*}{\textit{10 dB}} & \multirow{4}{*}{\textit{w/}}  & \multirow{2}{*}{\textit{Total}} & \multirow{2}{*}{\{\{472,525\},\{472,522\},\{434,504\}\}}          &  \multirow{2}{*}{2929}     \\
&&&&\\ \cline{3-5}
                       &                      &\multirow{2}{*}{\textit{Av.}}     & \multirow{2}{*}{\{\{2.36,2.63\},\{2.36,2.61\},\{2.17,2.52\}\}e-3} & \multirow{2}{*}{2.44e-3}  \\
&&&&\\ \cline{2-5}
                       & \multirow{4}{*}{\textit{w/o}} & \multirow{2}{*}{Total} &\multirow{2}{*}{\{\{1607,15\},\{1544,12\},\{1612,6\}\}}   & \multirow{2}{*}{4796}     \\
&&&&\\ \cline{3-5}
                       &                      & \multirow{2}{*}{\textit{Av.}}   & \multirow{2}{*}{\{\{8.04,7.5\},\{7.72,6\},\{8.06,3\}\}(e-3,e-5)} & \multirow{2}{*}{4e-3} \\
      &&&&\\                 \hline
\multirow{8}{*}{\textit{15 dB}} & \multirow{4}{*}{\textit{w/}}  & \multirow{2}{*}{\textit{Total}} & \multirow{2}{*}{\{\{440,894\},\{447,911\},\{440,894\}\} }      & \multirow{2}{*}{4026}     \\
&&&&\\ \cline{3-5}
                       &                      & \multirow{2}{*}{\textit{Av.}}   & \multirow{2}{*}{\{\{2.2,4.47\},\{2.24,4.56\},\{2.2,4.47\}\}e-4}  & \multirow{2}{*}{3.36e-4} \\
                       &&&&\\ \cline{2-5}
                       & \multirow{4}{*}{\textit{w/o}} & \multirow{2}{*}{\textit{Total}} & \multirow{2}{*}{\{\{2593,0\},\{2585,0\},\{2629,1\}\}}              & \multirow{2}{*}{7808}     \\
                       &&&&\\ \cline{3-5}
                       &                      & \multirow{2}{*}{\textit{Av.}}   & \multirow{2}{*}{\{\{1.3,0\},\{1.29,0\},\{1.31,5\}\}(e-3,e-7)}  & \multirow{2}{*}{6.51e-4} \\
                       &&&&\\ \hline
\end{tabular}
  \end{center}
  }
\end{table}

\section{Rate of Convergence} \label{sec:ROC}

The proposed \mbox{ad-hoc} DPCA increases and decreases the algorithmic complexity and \mbox{sum-rate} with marginal gaps,
respectively, while garnering \mbox{sub-stream} fairness thus improving BER. As shown before, our algorithm guarantees
feasible SINR targets via linear search. Moreover, the power control law used in the algorithm converges to a unique
fixed-point at a linear rate. For completeness, we review important definitions before presenting the proof, and begin with
the definition of linear convergence. A sequence $\{\xx^n\}\in\mathbb{R}^N$ converges to $\xx^\ast$ at a linear rate if there
exists a constant $c\in(0,1)$ satisfying
\begin{equation*}
\lim\limits_{n\rightarrow\infty}\frac{||\xx^n-\xx^\ast||}{||\xx^{n-1}-\xx^\ast||}=c,
\end{equation*}
where $||.||$ is some norm defined in $\mathbb{R}^N$. In a similar manner, for every initial vector $\xx^0$, the
sequence $\{\xx^n\}$ generated by an iterative algorithm converges to $\xx^\ast$ at a linear rate if
\begin{equation*}
||\xx^{n}-\xx^\ast||\leq c^n||\xx^{0}-\xx^\ast||
\end{equation*}
is satisfied, and $c\in[0,1)$. In other words, the distance $||\xx^{n}-\xx^\ast||$ is always lesser than
$c^n||\xx^{0}-\xx^\ast||$, and decays exponentially.

In \cite{225}, authors introduce contractive interference functions that do not need separate proof for the
existence of \mbox{fixed-points} and that also give estimates on the convergence rates of algorithms. These
generous features are not present in standard interference functions introduced in \cite{207}. Contractive
interference functions are obtained by slightly reformulating the last condition of standard interference
functions, scalability,
\begin{tabbing}
\= ~~ \textit{scalability:} $\forall\alpha>1,~\alpha\xtI(\xp)>\xtI(\alpha\xp),$
\end{tabbing}
while keeping the first two conditions, positivity and monotonicity, same. The last condition of contractive
interference functions, contractivity, in \cite{225} is given as
\begin{tabbing}
\= ~~ \textit{contractivity:} There exists a constant $c\in[0,~1)$ and a \\
\> ~~ vector $\xv>\mathbf{0}$ such that $\forall\epsilon>0,~ \xtI(\xp)+c\epsilon\xv\geq\xtI(\xp+\epsilon\xv)$.
\end{tabbing}
In \cite{225}, it is proven for contractive interference functions that for any initial power vector $\xp^0$,
the sequence $\xp^n=\xtI(\xp^{n-1})$ converges linearly to $\xp^\ast$
\begin{equation*}
||\xp^n-\xp^\ast||_\infty^{\xtv}\leq c^n||\xp^0-\xp^\ast||_\infty^{\xtv},
\end{equation*}
where $||.||_\infty^{\xtv}$ denotes weighted maximum norm for a given vector $\xv>\mathbf{0}$, i.e.,
$||\xx||_\infty^{\xtv}=\max_i|\frac{x_i}{v_i}|$. The reader is referred to \cite{225} for further details. Next
we show that interference function \eqref{eqn:ourSIF} is contractive.

\begin{figure}[htb]
\centering
\includegraphics[height=4.5cm, width=8cm] {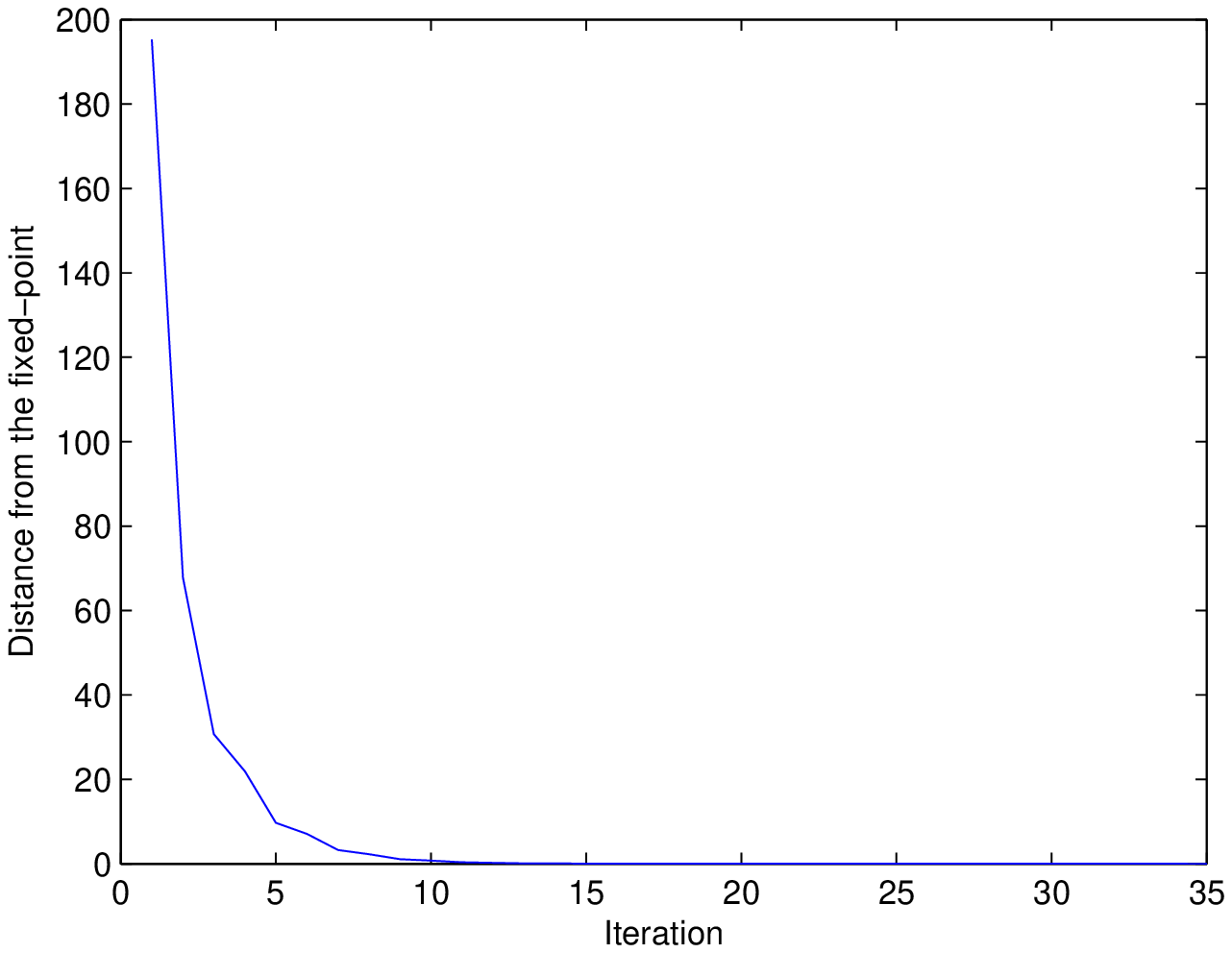}
 \caption{The distance between power vector and optimal point, $||\xp^n-\xp^\ast||_\infty$, decays at an exponential rate.} \label{Fig:Fig9}
\end{figure}

The interference function \eqref{eqn:ourSIF} can be rewritten as
$$\xtI\xmkl(\xp)=\sum_{d=1}^{d_j}\sum_{j=1}^{K}T_{k,j,d} p_{j,d}+N_k,$$
where $N_k=\frac{\Gamma_k}{G_{k,k}},$
\begin{equation}\label{eqn:entitites}
 T_{k,j,d} =
  \begin{cases}
   \frac{\Gamma_k G_{k,j,d}}{G_{k,k}} & \text{if } j\neq k, \\
   0       & \text{if } j=k,
  \end{cases}
\end{equation}
$G_{k,j,d}=|\xv\xmkl^\dagger\xH\xmkj\xu_{j,d}|^2$, and $G_{k,k}=|\xv\xmkl^\dagger\xH_{k,k}\xu_{k,l}|^2.$ Henceforth, the
contractivity condition is satisfied
\begin{eqnarray*}
  \xtI\xmkl(\xp+\epsilon\xv) &=& \xtI\xmkl(\xp)+\epsilon\sum_{d=1}^{d_j}\sum_{j=1}^{K}T_{k,j,d}v_j \\
   &\leq& \xtI\xmkl(\xp)+\epsilon\sum_{d=1}^{d_j}||T_d||_\infty^{\xtv}
v_k,
\end{eqnarray*}
where $T_d$ is a $K\times K$ matrix with entities in \eqref{eqn:entitites}. Thus, \eqref{eqn:ourSIF} is a
\textit{c}-contractive interference function with $c=\sum_{d=1}^{d_j}||T_d||_\infty^{\xtv}.$ The extension of
the above proof to interference function with min operator in Algorithm \ref{alg:AdHocAlg} is straightforward.
In Fig. \ref{Fig:Fig9}, the distance to a fixed-point $||\xp^n-\xp^\ast||_\infty$, the vector $\xv$ is chosen as
all ones vector (${\xv=\mathbf{1}}$) thus omitted in notation, is plotted for a channel realization at SNR 30
dB.

\section{Conclusion}

We developed an \mbox{ad-hoc} DPCA that achieves \mbox{sub-stream} fairness at the cost of a slightly increased algorithmic
load. The instantaneous \mbox{sum-rate} degradation of the proposed algorithm is lesser since \mbox{sub-stream} fairness poses
milder conditions than stream fairness. The proposed algorithm can achieve stream fairness in the ergodic sense as well. The
algorithm guarantees feasible SINR targets via linear search. As opposed to common approach, the algorithm does not require
preset SINR targets, and optimizes the targets. In the paper, the impacts of \mbox{sub-stream} fairness on BER results and
algorithmic parameters on benchmarking processes are illustrated. Finally, via contractive interference functions, the power
control law in the proposed algorithm is proven to linearly converge to a unique fixed-point.

In addition to the future research directions already pointed in the paper, theoretical modeling of numerical BER results
presented in the paper is another important research direction. Finally, as the system size increases, e.g., increasing number
of users and \mbox{sub-streams}, achieving theoretically promised results in practice becomes a challenging research problem,
especially in the high SNR regime since by and large, algorithms require more iterations in this regime. Therefore, developing
fast converging beamforming algorithms is consequential.

\bibliographystyle{IEEEtran}
\bibliography{IEEEabrv,IEEEfull}


\end{document}